\documentclass[12pt]{article}
\usepackage{cite,dsfont}
\usepackage{amsfonts}
\usepackage{amssymb}
\usepackage{amsmath}
\usepackage{graphicx}
\usepackage{slashed}
\usepackage{setspace}

\usepackage{color}

\newcommand{\cn}{{\cal N}}

\newcommand{\reef}[1]{(\ref{#1})}

\newcommand{\ca}{{\cal A}}

\newcommand{\cl}{{\cal L}}

\newcommand{\co}{{\cal O}}

\newcommand{\be}{\begin{equation}}
\newcommand{\ee}{\end{equation}}

\def\be{\begin{equation}}
\def\ee{\end{equation}}
\def\bea{\begin{eqnarray}}
\def\eea{\end{eqnarray}}
\def\ba{\begin{array}}
\def\ea{\end{array}}
\def\bd{\begin{displaymath}}
\def\ed{\end{displaymath}}

\def\a{\alpha}
\def\b{\beta}

\def\d{\delta}
\def\e{\epsilon}           
  
\def\g{\gamma}

\def\l{\lambda}
\def\m{\mu}
\def\n{\nu}
\def\o{\omega}  
\def\r{\rho}                                     
\def\s{\sigma}                                   

\def\D{\Delta}

\def\L{\Lambda}

\def\pa{\partial}                              
\def\>{\rangle} 
\def\<{\langle} 
\def\Dsl{D \hskip-.6em \raise1pt\hbox{$ / $ } }
\def\to{\rightarrow}
\def\pa{\partial}

\def\lab{\label}

\newcommand{\eps}{\epsilon}

\def\bz{\bar z}
\def\bF{\bar F}

\begin{document}
\begin{titlepage}

\begin{flushright}
\texttt{MITY-CTP/4879}\\
\end{flushright}

\begin{center}
{\Large \bf Lecture for the Fortieth Anniversary of Supergravity\\} 

\vspace*{0.5 cm}

{\bf Daniel Z.~Freedman,$^{1,2}$ }
\bigskip

$^{1}$Stanford Institute for Theoretical Physics and Department of Physics\\
Stanford University\\ Stanford, CA 94305, USA
\bigskip

$^2$Department of Mathematics and Center for Theoretical Physics\\
 Massachusetts Institute of Technology\\
Cambridge, MA 02139, USA
\bigskip

\texttt{dzf@math.mit.edu} \\
\end{center}

\vspace*{0.25cm}
\begin{abstract} 
{}
\noindent{}   In the first part of this lecture, some very basic ideas in supersymmetry and supergravity are presented at a level accessible to readers with modest background in quantum field theory and general relativity.  The second part is an outline of a recent paper of the author and his collaborators on the AdS/CFT correspondence applied to the ABJM gauge theory with $\cn =8$ supersymmetry.

\end{abstract}

\end{titlepage}

\section{Introduction:}

The first paper on supergravity in $D=4$ spacetime  dimensions  \cite{SG1} was submitted to the Physical Review in late March, 1976.  It was a great honor for me that the fortieth anniversary of this event was one of the features of the 54th Course at the Ettore Majorana Foundation and Centre for Scientific Culture in June, 1976.  This note contains some of the material from my lectures there. The first part focuses on the most basic ideas of the subjects of supersymmetry and supergravity, ideas which I hope will be interesting for aspiring physics students. The second part summarizes the results of the paper \cite{FPPW} on what might be called a curiosity of the AdS/CFT correspondence. 

\section{Fundamentals of SUSY and SG}
Supergravity is built upon the principle of supersymmetry.  Global supersymmetry was discovered in the early 1970's, see \cite{susy1, susy2,susy3}, and rapid development then followed.   The theory postulates a symmetry between fermions and bosons.  The symmetry is generated by  conserved charges that are spacetime spinors $Q_\a,~ \a=1,2,3,4$.  Since they are spinors, it is natural that they enjoy simple properties under anti-commutation, and the basic supersymmetry algebra is given by
\be \lab{susyalg}
\{Q_\a,\bar Q^\b\} = - (\g_\m)_\a{}^\b P^\m,
\ee   
where $P^\m,~\m=0,1,2,3$ are the generators of spacetime translations. 
In a supersymmetric theory the fields and the particles they describe are grouped in multiplets which contain both bosons and fermions.  Supersymmetry is the only framework that permits the unification of particles of different spins within a single algebraic structure, called a superalgebra.  

The parameters of {\it global} SUSY transformations are {\it constant} spinors $\e_\a$.  The appearance of translation generators $P^\m$
on the right side of \reef{susyalg} suggests a relation to the geometry of spacetime.  From this one can envisage a supersymmetric theory of gravity in which the graviton acquires a fermionic partner, the gravitino.
In a gravitational theory the constant spinors must be replaced by arbitrary functions of $\e_\a(x)$ of the spacetime coordinates.  In this way supergravity becomes the gauge theory of supersymmetry.

I  read the path-breaking paper of Wess and Zumino \cite{susy3} when it first appeared in early 1974. I was deeply impressed, and no other single paper has so transformed my subsequent professional life. The idea that there is a consistent symmetry principle that relates particles of different spin and statistics came as a thunderbolt.  
SUSY and SG are attractive to me because the component formalism makes close contact with the basic principles of quantum field theory and their close relation with geometry.  These principles include:\\
\begin{enumerate} 
\item Relativistic treatment of spin via Dirac $\g$-matrices.  By this I mean the complete set
\be
{I,~ \g^\m,~ \g^{\m\n},~ \g^{\m\n\r}~\ldots}, 
\ee
and their algebraic relations, e.g. { $\g^\m\g^{\n\r} = \g^{\m\n\r} + \eta^{\m\n} \g^\r-\eta^{\m\r}\g^\n$}. Note that these matrices of the Dirac-Clifford basis are totally antisymmetric in their tensor indices.

\item Basic equations of the geometry of gauge fields and Riemannian geometry,  e.g.  gauge field Bianchi identity
\be
{ \pa_\m F_{\n\r} +\pa_\n F_{\r\m}+\pa_\r F_{\m\n}=0.}
\ee

\item Spin and statistics:  { $\{\psi_\a(x),\psi_\b(y)\} =0$} and $\{\e_\a,\e_\b\} =0$  even at the level of classical manipulations.

\end{enumerate}

As the first example of the interplay of these principles, let's consider the action of the 
free Maxwell-Dirac theory:  
{ \be
S = -\int d^Dx [\frac14 F^{\m\n}F_{\m\n} + \bar\psi\g^\m\pa_\m\psi]
\ee}
We will show that it is invariant under SUSY trf. rules in any spacetime dimension $D$.:
{ \be
\d A_\m = -\bar\eps \g_\m\psi +\bar\psi\g_\m\eps,\quad\qquad \d\psi =\frac12 \g^{\n\r}F_{\n\r}\eps,\qquad\quad \d\bar\psi = - \frac12 \bar\e \g^{\n\r}F_{\n\r}.
\ee}
In the proof it suffices to consider only the terms containing $\eps$, since the terms with $\bar\eps$ are related by complex conjugation.  The separate variations of the Maxwell and Dirac actions are 
{\bea
\d S_1 &=& -\int d^Dx F^{\m\n}\pa_\m \d A_\n=+\int d^Dx \,\pa_\m F^{\m\n}\d A_\n\\
&&\rightarrow_{SUSY}\int d^Dx \,\pa_\m F^{\m\n}\,(\bar\psi\g_\n\eps)\\
\d S_{1/2} &\rightarrow_{SUSY}&-\frac12 \int d^Dx \,(\bar\psi \g^\m \g^{\n\r}\eps)\, \pa_\m F_{\n\r}
\eea}
We need to show that the sum vanishes.  This requires the $\g$-matrix algebra relation given above which we insert in $\d S_{1/2}$:

{\bea 
\d S_1 &=& \int d^Dx \,\pa_\m F^{\m\n}\,\bar\psi\g_\n\eps\\
d S_{1/2} &=&-\frac12 \int d^Dx \,(\bar\psi[ \g^{\m\n\r} + \eta^{\m\n} \g^\r-\eta^{\m\r}\g^\n]\eps)\, \pa_\m F_{\n\r}
\eea}

The last two terms cancel with ${ \d S_1}$, so first term must cancel by itself.  Indeed, it does cancel using the gauge field Bianchi identity which is repeated from above for full clarity:

$${ \g^{\m\n\r}\pa_\m F_{\n\r} =\frac13 \,\g^{\m\n\r}(\pa_\m F_{\n\r}+\pa_\m F_{\n\r}+\pa_\m F_{\n\r}) = 0}.$$

The relation between spin and statistics played no role in this proof because we dealt with a {\it free} theory. In {\it interacting} SUSY theories it enters in a big way. For example, the proof of invariance of any interacting SUSY requires a manipulation of cubic terms in the fermions known as a Fierz rearrangement. Fierzing can be tricky, so I will just say that the basic Fierz relation is based on the completeness of the full set of Dirac $\g$-matrices  in the particular spacetime dimension under study. 

Actually, the basic spinor field in 4 spacetime dimensions is not a complex Dirac spinor $\psi_\a(x)$. Rather it is a Majorana  spinor\footnote{One key discovery of Ettore Majorana was that in spacetime dimension  $D=4$ the basic fermion field  $\l_\a(x)$ (and the Dirac $\g_\m$ matrices) can be chosen to be real.  The chiral projections $L\l = \tfrac12(I+\g_5)\l$ and $R\l = \tfrac12(I-\g_5)\l$ are effectively two-component Weyl spinors. Supersymmetric theories can be formulated using either Majorana or Weyl spinors.} $\l_\a(x).$
Suppose we assume that the components of $\l_\a(x)$ are ordinary numbers, i.e. $  [\l_\a(x),\l_\b(y)] =0.$
\,One can show using the reality property that the basic free spinor kinetic and mass terms vanish:
\be
S_{\text Majorana} = -\frac12\int d^4 x\bar\l (\g^\m\pa_\m- m)\l=-\frac14\int d^4 x\pa_\m(\bar\l \g^\m\l)=0,
\ee
so a free Majorana theory of the wrong statistics is vacuous.

Let's move on to something more sophisticated. In the (highly recommended) book \cite{book}, it is called "the universal part of supergravity." It comprises the first steps in the proof of invariance of local supersymmetry which are valid in any spacetime dimension and for any type of spinor.  It is my dream that these steps one day become a standard part of the curriculum in quantum field theory courses.  

In supergravity the fermionic partner of the spin 2 graviton is the spin 3/2 gravitino. This is described by a spinor-vector field $\psi_\mu(x)$, whose suppressed spinor index $\a$ takes $2^{[D/2]}$ values in $D$ dimensions.  This field was first considered  by Rarita and Schwinger in 1941.  When  spinors are present in a gravity theory, the basic bosonic variable becomes the frame field $e^a_\m(x)$. It is related to the metric tensor by $g_{\m\n} =e^a_\m \eta_{ab}e^b_\n$, where $\eta_{ab}$ is the Minkowski metric of flat spacetime. I must assume that readers have some basic knowledge of the way connections and curvatures work in the frame field formalism. (If not they can consult \cite{book}.)

We begin by writing the action of the universal part of supergravity.  It is the sum of the standard Hilbert action $S_2$ plus the massless gravitino action $S_{3/2}$.  Note that this involves the third rank element $\g^{\m\n\r}$ of the Dirac algebra. 
{\bea 
S_2 &=& \frac12\int d^Dx \,e\, e^{a\m}e^{b\n}R_{\m\n ab} =\frac12\int d^Dx \sqrt{-g} \,g^{\m\n}R_{\m\n}\\
S_{3/2} &=& -\int d^Dx \,e \,\bar\psi_\m \g^{\m\n\r}D_\n\psi_\r\\
&& D_\n\psi_\r \equiv (\pa_\n+\frac14\o_{\n ab}\g^{ab})\psi_\r
\eea}
We need the local SUSY variations of these fields. For the gravitino we take
\be
\d\psi_\r = D_\r\eps(x)=(\pa_\r+\frac14\o_{\r ab}\g^{ab})\e(x), \qquad\quad \d\psi_\r = \bar\eps(x)\stackrel{\leftarrow}{D}_\r.
\ee
This bears a close resemblance to ${ \d A_\r  =\pa_\r \theta(x)}$ in Maxwell gauge theory and its non-abelian extension.  Indeed one of the good things that happens in supergravity is that it contains a new (in 1976) spin 3/2 gauge principle which joins the spin 1 gauge principle of electromagnetism and Yang-Mills theory and the spin 2 gauge principle of gravitation.  These are the only gauge theories which allow consistent interactions.\footnote{Vasiliev theories which involve an infinite number of higher spin gauge fields are also consistent \cite{Vasiliev:1990en}.}  The frame field and metric variations are
\be
\d e^a_\m =-\frac12 \bar\psi_\m \g^a\eps + c.c. \implies\d g^{\m\n} = \frac12(\bar\psi^\m \g^\n +\bar\psi^\n\g^\m)\eps+c.c.
\ee
We assume that the gravitino  field is complex because that case should be more familiar for readers.
In our proof that ${ \d(S_2 +S_{3/2})}$ vanishes to \emph{lowest order} in ${ \psi}$,  it is sufficient to study terms containing $\e$ and $\bar\psi$.  If these terms vanish,  so do the conjugate terms.

Let's start with $S_2$ and write its standard Euler variation
{\bea   \lab{dels2}
\d S_2 &=& \frac12\int d^Dx 
 \sqrt{-g}\,\d g^{\m\n}\,(R_{\m\n} -\frac12 g_{\m\n}R)\nonumber\\
&=&\frac12\int d^Dx \sqrt{-g}\,[(\bar\psi^\m\g^\n\eps)\,(R_{\m\n} -\frac12 g_{\m\n}R) +c.c.]\lab{dels2}
\eea}
For $S_{3/2}$ we permute indices, $\m\n\r\to \r\m\n$, insert $\d \psi_\n$,  and note that the commutator of covariant derivatives gives us the Riemann curvature tensor.   This relation is called the Ricci identity of differential geometry. 
{\bea
\d S_{3/2} &=& -\frac12 \int d^Dx \sqrt{-g}\bar\psi_\r\g^{\r\m\n}[D_\m, D_\n]\eps\nonumber\\
&=&-\frac18\int d^Dx \sqrt{-g}\bar\psi_\r\g^{\r\m\n}\g^{ab}R_{\m\n ab}\eps.\lab{dels32}
\eea}
Somehow we have to show that the spinor bilinear with the product of five $\g$-matrices in \reef{dels32}simplifies and then cancels the single  $\g$-bilinear in \reef{dels2}.  This involves
$\g-$algebra that is more difficult than previous but doable with some practice.  The product $\g^{\r\m\n}\g^{ab}$ can be expressed uniquely as the sum of 5th rank, 3rd rank, and 1st rank $\g$'s:
{\bea
\g^{\r\m\n}\g^{ab}R_{\m\n ab}&=& [\g^{(5)}+\g^{(3)}+\g^{(1)}]R_{....}\\
&=& \g^{\r\m\n ab} R_{\m\n ab} + 2\g^{\m\n b}R_{\m\n{}\r{}b}+4 \g^{\n\r b}R_{\m\n}{}^\m{}_b\\
&+&  4 \g^\m R_{\m\n}{}^{\r\n} - 2\g^\r R_{\m\n}{}^{\m\n}\,.
\eea}
The 5th rank $\g$-matrix vanishes in 4 dimensions. The matrix is present for $D\geq 5$, but there is a cyclic contraction of indices which produces the first Bianchi identity of the curvature tensor, so the 5th rank term vanishes.  The first rank 3 term also vanishes by the Bianchi identity, and second one vanishes because 
$\g^{\r\n\s}$ is anti-symmetric, while the Ricci tensor $R_{\n\s}$ is symmetric.\footnote{We converted from frame indices to coordinate indices to make this statement, e.g.  $\g^\s = \g^b e_b^\s$.}  We are left with the rank 1 terms in the last line.  It is easy to substitute them in \reef{dels32} and find that the sum $\d S_{3/2} +\d S_2 =0$,  which establishes the result we wanted.

The deed is done!  I hope that readers noticed how Dirac algebra and Riemannian geometry conspired to make it happen. This is the universal part of supergravity, but it just the first step of a complete proof. For example, in the simplest case of pure $\cn=1,~D=4$ supergravity, there are order $\psi^3\e$ terms that arise
from the variation $\d \g^\mu =  \g^a \d e_a^\m =-\tfrac12 \g^a(\bar\psi_a\g^\m\e + c.c.)$.  Order $\psi^5\e$ terms also arose by the techniques used in \cite{SG1}, but we have learned how to avoid them \cite{SG2, SG3,SG4}.  Cancellation of such higher order terms is a delicate business.  

It is even more delicate in extended SG theories in $D=4$ dimensions in which the graviton has $\cn\geq 2$ gravitino partners and lower spin fields are also needed to complete the bose-fermi supermultiplet.  Sophisticated calculations are also needed to construct SG theories for $D\geq 5.$  It turns out that  extended SG theories exist up to $\cn=8$ in $D=4$, and there is also a limit $D = 11$ for higher dimensional  SG.  The two maximal theories have been constructed in \cite{Neq8} and \cite{SGD11} and  are closely related.  The $D=11$ theory can be dimensionally reduced on the 7-dimensional torus, and this process gives $\cn=8,~D=4$ supergravity.  The limits $\cn =8$ and $D=11$ are related to the remarks above concerning consistent gauge principles. The supermultiplets needed to exceed $\cn =8$ and $D=11$ necessarily contain higher spin gauge fields such as the symmetric tensor spinor $\psi_{\m\n}(x)$, and there are no consistent interacting theories for such fields. 

I would like to close this section with some brief comments about what has been achieved within the framework  of supersymmetry and supergravity and what has not been achieved.  What has not yet been achieved is clear experimental support;  the necessary superpartners of the known elementary particles have not been found up to the scale $M \approx 1-2$ TeV accessible at the LHC accelerator.  Arguments based on the idea of "naturalness" of electroweak symmetry breaking  suggested that evidence for SUSY would be found at the LHC, but  Nature apparently does not know about this form of "naturalness."

What has been achieved is a vast theoretical framework which has greatly influenced modern theory.  Global supersymmetry  in $D=4$ gives significant control of strong coupling effects \cite{SW}.
The maximal $\cn =4, D=4$ global theory \cite{Neq4} is ultraviolet finite. It is a superconformal theory.  Techniques have been developed \cite{BCFW} to calculate its on-shell amplitudes by methods that are far simpler than the evaluation of many Feynman diagrams. 

The study of $\cn =2,~D=4$ supergravity has led to the development of "special geometry" of considerable influence in both physics and differential geometry \cite{specgeom}.  Theorists have speculated that the maximal $\cn=8,~D=4$ SG theory is an ultraviolet finite theory of quantum gravity.  This idea has passed calculation tests up to 4-loop order \cite{Bern:2009kd}.  The on-shell amplitudes program \cite{EH} can be applied to this theory and results \cite{howe, beisert} show that  divergences cancel through 6-loop order, but cannot be excluded beyond this level. If the theory is finite to all orders, a new principle that goes beyond the obvious symmetries must be found.  Another important line of investigation concerns black hole solutions of supergravity theories \cite{HorStr}.  One interesting feature is the attractor mechanism\cite{attract} for solutions in theories with scalars.
Finally there is a close relation between 10- and 11-dimensional supergravity and superstring theory and M-theory.  Roughly speaking supergravity is the low energy limit of superstring theory, and many papers that address ideas in string theory actually work at the supergravity level.  There are D-brane solutions in supergravity, and much of the calculational support for the AdS/CFT correspondence comes from calculations that compare quantities in  maximal $\cn=8,~ D=5$ SG \cite{GRW} with the   $\cn =4, D=4$ superconformal theory.

The comments above are far too brief to cover this vast area of theoretical work. References are meant to be suggestive and are far from complete. I was told a few months ago that there are about 14,000 papers on the SPIRES archive with the word "supergravity" in the title or as a keyword.\footnote{Martin Rocek, private communication.}

\section{Boundary interactions in the AdS/CFT correspondence}

The AdS/CFT correspondence was invented \cite{juan} by Juan Maldacena in late 1997.  At the time of the Erice program in June 2016, his paper had 11,984 citations. At the time of this write-up, 6 months later, the number has increased to 12,449 !!!. These numbers prove that AdS/CFT permeates fundamental theoretical physics!

The basic assertion of AdS/CFT is that the observables in two very different types of field theories should be equal. 
On the CFT side, there is a conformal invariant gauge theory in $d$ spacetime dimensions and NO gravity.   The observables are correlation functions of gauge invariant composite operators:  $\<\co_1(x_1)\co_2(x_2)\ldots\co_n(x_n)\>.$ The $x_i$ are points in flat spacetime. It is difficult to calculate these correlators because strong coupling methods in field theory are often needed and are usually very crude.  An exception occurs in a few SCFTs where the method of supersymmetric localization can be used
to calculate some important observables such as the free energy.

The AdS side is a gravity or SG theory in $d+1$-dimensional spacetime.  That theory must have a special solution called anti-de Sitter space.  AdS is a spacetime with a boundary.  AdS/CFT asserts that the boundary limits of the bulk fields act as sources for the field theory operators.
It is remarkable that computations of correlators from the gravity theory involve considerably simpler classical techniques. One must solve partial differential equations and do some integrals. 
These techniques were initially developed by  Gubser, Klebanov, Polyakov \cite{GKP} and by Witten \cite{Witten}.  Their papers have  6789 ($\to$ 7063) and 7857 ($\to$ 8181) citations.  Many theorists have contributed to the subsequent development.
  
I will discuss an interesting feature of the  AdS/CFT correspondence applied to the duality between the
$\cn =8, ~d=3$  ABJM ~CFT and $ \cn=8,~D=4$  SG.  My paper on this topic, written with 
 S. Pufu, K. Pilch and N. Warner \cite{FPPW}, was submitted to the archive well after the Erice program.
In this note I will summarize the ideas  and results; perhaps this will motivate some readers to look at the paper.
The models on both sides of the duality are invariant under the symmetry group SO(8). So operators in the CFT and fields in the gravity dual live in representations of SO(8).  The conformal group in 3 spacetime dimensions is SO(3,2) and this is also the isometry group of AdS$_4$. The conformal symmetry implies that operators are classified by their scale dimension $\D$ and spin $s$.
Of course, the operators and fields of different spin are also organized in supermultiplets. A supermulitplet is a representation of the conformal superalgebra OSp(8,4).

The ABJM theory \cite{ABJM} contains $\D=1$ real scalar operators $\co_{IJ}(x)$ in the $35_v$ representation of SO(8). 
These operators have a non-vanishing 3-point correlator
 {$\<\co_{IJ}(x)\co_{KL}(y)\co_{MN}(z)\>\ne 0.$}. This can be  calculated  exactly in the CFT because $\co_{IJ}(x)$ is in a short supermultiplet  whose top component is $T_{\m\n}$.  This allows  application of the method of supersymmetric localization \cite{pestun}.
 
 The standard method of calculation \cite{9804058}  of 3-point correlators in the gravity dual of a CFT is by evaluation of a Witten diagram containing a cubic coupling from the bulk Lagrangian. Gauged $\cn=8,~D=4 ~SG$ contains 35 fields $A^{IJ}$ dual to the $\co_{IJ}$, \emph{ but  there is no cubic $A^3$ coupling}.  Something new must be found to produce $\<OOO\>$ from bulk SG!  This is the rather acute puzzle that was solved in \cite{FPPW}.

The resolution of this puzzle is that  SUSY requires that renormalized on-shell bulk action contains a {\emph cubic BOUNDARY term} in addition to standard boundary terms from holographic renormalization. The
new boundary term is
\be
S_3  =\frac{1}{8\pi G_4}\frac16 \int d^3x\sqrt{-h}\, A^{IJ}A^{JK}A^{KI}
\ee
\noindent This boundary term does produce $\<\co_{IJ}(x)\co_{KL}(y)\co_{MN}(z)\>$, and it
matches the CFT result.

It is worth contrasting the situation in ABJM theory with that of  four-dimensional \hbox{$\cn=4$} supersymmetric Yang-Mills theory.  In the latter theory the basic  chiral primary operators $\co_{\Delta=2}$ are also in the same short multiplet as the stress tensor.  But their 3-point
 correlators    
are \emph{protected}  \cite{Lee,DFS}
This means that they are independent of the gauge coupling constant. They can be computed at weak coupling by performing Wick contractions of free fields and these results match the strong coupling calculations in the gravity dual.  This is \emph{not} true for the scalars $\co_{IJ}$ of  $\cn=8$ ABJM theory, where there are strong coupling effects.
So the agreement between
the gravity and gauge theory results in ABJM is a \emph {precision test of holography.}

\subsection{A consistent truncation of the bulk supergravity theory}

To avoid dealing with fields with many indices, we now introduce a consistent $\cn=1$ truncation of the bulk $\cn=8,~D=4$ SG theory that faithfully captures the dynamical content of interest.  In fact, this truncation was studied earlier \cite{FP} and it was shown that  the free energy in the gravity dual precisely matches a deformation of ABJM in which conformal symmetry is broken by mass terms.  The truncated theory contains
the gravity multiplet $e^a_\m,~\psi_\mu$ coupled to three chiral multiplets,  $z^\a=A^\a +iB^\a,~\chi^\a,~\a =1,2,3$.   We need only the classical bosonic action
{\bea \lab{Sbosonic}
S =  \frac{1}{8 \pi G_4} \int d^4x \sqrt{-g} \bigg[ \frac 12 R - \sum_{\alpha = 1}^3\frac{ |\pa_\m z^\alpha|^2 }{(1 - |z^\alpha|^2 )^2} 
   +\quad \frac{1}{L^2} \left(- 3 + \sum_{\alpha = 1}^3 \frac{2}{1 - |z^\alpha|^2}\right)\bigg].
\eea} 
The overall normalization is given by Newton's constant $G_4$, and $L$ is the length scale of AdS.  This is a quite simple theory.  The scalar kinetic term describes a nonlinear $\s$-model in which the target space is a K\"ahler manifold that consists of three decoupled copies of the Poincar\'e disc. The scalar potential also consists of three decoupled terms.
It is clear that the Lagrangian contains NO CUBIC TERMS.  Hence the puzzle described above.

In $\cn =1$ SG,  the scalar potential $V(z,\bar z )$  is related to the holomorphic superpotential by the quadratic formula:
\be
V = e^K[\nabla_\a W\, K^{\a\bar\b}\nabla_{\bar\b}\bar W -3 W\bar W]\qquad\quad \nabla_\a W \equiv (\pa_\a + K_\a)W,
\ee
where $K$ is K\"ahler potential
\be
K = -\sum_{\a=1}^3 \ln(1-\bar z^\a \bar z^\a)
\ee
and $K_\a = \pa K$.  The result is $ W= (1 + z^1z^2z^3)/L.$  It is an algebraic miracle that a highly coupled $W(z)$ corresponds to a completely uncoupled  $V(z, \bar z)$!  The cubic term in $W(z)$ turns out be exactly what we need to produce the problematic 3-point correlation function, but we need to move it into the action, because it is the on-shell bulk action that is the generating functional of CFT correlators in AdS/CFT.  

We gave two arguments for this in \cite{FPPW}.  The first, actually  done first in \cite{FP}, was by a Bogomolny analysis, which I will not repeat here.  The second method, which is more rigorous, was to extend local supersymmetry to the AdS boundary, as we now motivate.
i. In the usual proofs of invariance in SG, one is happy to show that the variation of the action reduces to a total spacetime derivative, i.e.   $$ \d S =\int d^4x\pa_\mu [\sqrt{-g}\,\bar\eps(x) X^\m],$$ where $X^\m$ is vector-spinor function of the fields of the theory.  This is correct for most purposes, since the spinor parameters $\e_\a(x)$ are arbitrary functions which can be assumed to vanish at large distance.  \\
ii. However, in AdS/CFT the behavior as $r\to\infty$ is crucial,  where $r$ is the radial coordinate of the AdS metric, and the boundary is reached as $r\to \infty.$  The $\eps(r,x)$ are Killing spinors and the fields vanish at rates fixed by field eqtns.  So  we collect  bdy terms and write
{ $$ \int d^4x\pa_\mu [\sqrt{-g}\,\bar\eps(x) X^\m] = \int_{r=r_0}d^3x\sqrt{-h}\, \bar\eps\, X^r \equiv \d S_{bdy}.$$} \noindent Here $r_0$ is a cutoff which we eventually take to infinity.\\
iii. The final step is to find a set of counterterms { $S_{CT} = \int d^3x \sqrt{-h}\, \cl_{CT}$,} whose SUSY variation cancels the boundary variation, i.e.
 $$ \d_{SUSY} S_{CT} = - \d S_{bdy}.$$ 

It is simple and instructive to work out boundary terms and the counterterms that cancel them in the global limit of a general $\cn=1$ SG model in AdS spacetime.  This is a limit in which the back reaction of the matter fields is consistently suppressed.  The result is an action that has global SUSY on AdS$_4$ and is 
similar to the construction of \cite{FS}.\\
a.  In this global limit, the SUSY parameters are AdS Killing spinors. Killing spinors satisfy 
{$$ (D_\m +\frac{1}{2L}\g_\m)\eps(r,x) =0 $$} 
They can be found explicitly \cite{book} for the AdS$_4$ metric  { $ ds^2 = dr^2 + e^{2r/L} \eta_{ij} dx^i dx^j , i,j =0,1,2 $.}   Their leading components  grow at the boundary. as \quad {$\eps(r,x)\sim e^{r/2L}.$}\\
b.  This limiting  procedure works for any K\"ahler metric and any superpotential of the form
{$ W = (1 + W(z^\a))/L $} with cubic $W(z^\a)$ .  This guarantees that the SG model has an AdS stationary point with cosmological constant $\L = -3/L^2$,  the SUSY value.  \\
c.  There are further simplifications;  the information on the counterterms that we need is captured by the case of one chiral multiplet { $z,~\chi$} with a flat K\"ahler potential { $ K = z \bz$} and cubic { $ W = z^3/3$}.
The result in this simple case extends immediately to the three field truncation with  $W= z^1z^2 z^3 $.\\
d. The  result is a simple  action\footnote{The chiral projector is $P_L=(1+\g_5)/2.$} with auxiliary fields $F,~\bar F$.
{ \bea
S &=& S_{kin} + S_F + S_{\bar F}\\
S_{kin} &=& \int d^4x\sqrt{-g}\bigg[- \pa_\m z\pa^\m\bz - \frac12 \bar\chi \g^\m D_\m\chi \\
&&\quad+ (F+ z/L)(\bar F + \bar z/L) + 2 z\bz/L^2 \bigg]\\
S_F &=& \int d^4x \sqrt{-g}[ F W'- \frac12 W"\bar\chi P_L\chi + 3W/L]\\
S_{\bar F} &=&  (S_F)^*\,.
\eea}
The 3 terms { $ S_{kin},~S_F,~ S_{\bar F}$} are \emph {separately} invariant under:
{ $$  \d z = \bar\eps P_L\chi\qquad \d P_L \chi = P_L(\g^\m\pa_\m z +F)\eps \qquad \d F = \bar\eps(\g^\m D_\m - 1/L)P_L\chi .$$}  
$S_F$ is very simple and so is its SUSY variation.  It vanishes in flat spacetime, and the remaining AdS terms give
{ \be
\d S_F = \int d^4x \sqrt{-g}[\nabla_\m(\bar\eps \g^\m W' P_L\chi) -\bar\eps( \stackrel{\leftarrow}{D}_\m\g^\m -2/L)W'P_L\chi ].
\ee} 
The last term vanishes by the (adjoint of the) Killing spinor equation.\\
The first term is the boundary term we are looking for!  It is cancelled by the counterterm 
{\be
S_{CT} = -\int d^3x \sqrt{-g}\, W(z).
\ee}
The same argument applied to $S_{\bF}$ gives the complex conjugate term.  

We restore the normalization and choose $W= z^1z^2z^3/L$ for the three-field model.  This gives the net cubic counterterm
\be\lab{bdyct}
S_3 = -\frac{1}{8\pi G_4L}\int d^3x \sqrt{-g} (z^1z^2z^3 + c.c.)\,.
\ee
Let's examine the behavior as the cutoff $r_0\to\infty.$ In the coordinates we are using, $\sqrt{-h}=\sqrt{-g} = e^{3r_0/L}$.  The scalar fields  $A^\a = \text{Re} \,z^\a$ approach the boundary at the rate $e^{-r_0/L}$. So this counterterm is finite, while counterterms from  the holographic renormalization procedure diverge at the boundary.  Those counterterms would be obtained from $\d S_{kin}$.  See \cite{FPPW} for a more complete discussion.

\subsection{Alternate Quantization}

The counterterm $S_3$ is exactly what we need to calculate the correlator $\<\co_1(x) \co_2(y)\co_3(z)\>$,
but the pathway to get there is rather long. The reason for this is that
 \emph{alternate quantization} must be applied to a scalar field which sources an operator with $\D=1$ in a $d=3$ dimensional CFT.  To introduce this topic, let's consider the 
two branches of the AdS/CFT mass formula for scalar fields
{ $$\D=(d \pm\sqrt{d^2 +4 m^2L^2})/2 .$$}
We also note that the general solution of the free scalar equation of motion in AdS$_{d+1}$  has two distinct
asymptotic  behaviors
$$ A(r,x) = A_1(x) e^{(\D-d)r/L} + A_2(x)e^{-\D r/L}.$$
Most applications of AdS/CFT require only the upper branch in which the bulk scalars source operators with $\D > d/2$.  In this case  the source is always the  coefficient $A_1(x)$ of the leading asymptotic term.  
However, for operators in the range\footnote{The lower limit is the unitarity bound. Note that a {\it free} field operator in a CFT$_d$ has $\D_{\text{free}} = (d-2)/2.$} $(d-2)/2 < \D <d/2$, the situation is very different.
The source is the Legendre transform \cite{KW} of the on-shell bulk action with respect to $A_1(x)$. 

The procedure has several steps which we summarize here for a model with a single scalar field  $A(r,x)$ with  the cubic boundary \reef{bdyct}.
The Legendre transform, which replaces the on-shell action as the generating functional of correlation functions, is defined as
 \be
  \tilde S_\text{on-shell}[\ca] = S_\text{on-shell}[A_1] + \int d^3x \,  \ca(x) A_1(x)  \,
 \ee 
This must be extremized  with respect to $A_1(x)$ to obtain $\ca(x)$ as a non-local functional of $A_1(x)$, and this relation is then inverted to find $A_1[\ca]$. The result is 
\be
  A_1(x) = -\int d^3 y \, \frac{\ca (y)}{2 \pi^2 |x-y|^2} - \frac{1}{(2 \pi^2)^3} \int d^3 y\, d^3 z\, \ca (y) \ca (z) I(x, y, z) \,.
\ee 
The important quantity for us is the conformal integral
 \be
 I(x, y, z) = \int d^3 w \, \frac{1}{|x-w|^2 |y-w|^2 |z-w|^2} = \frac{\pi^3}{|x-y| |y-z||x-z|} \,
 \ee 
which was evaluated using the method of conformal inversion \cite{9804058}.  Finally we can reassemble things and write
\bea
  \tilde S_\text{on-shell}[\ca ] &=& -\frac 1{4 \pi^2} \int d^3x \, d^3 y\, \frac{\ca (x) \ca(y)}{ |x-y|^2}\\
   &-& \frac{1}{24 \pi^3} \int d^3 x\, d^3 y\, d^3 z \frac{\ca(x) \ca(y) \ca(z)}{ |x-y| |y-z| |x-z|}  + O(\ca ^4) \,.
\eea
This is the generating functional; one applies functional derivatives with respect to $\ca(x)$ to obtain  correlators.  We restore the normalization and write the results for the truncation to three chiral multiplets:
 \bea \lab{3pttrunc}
   \langle {\cal O}_\alpha(\vec{x}_1) {\cal O}_\beta(\vec{x}_2) \rangle &=& \frac{ L^2}{2 \pi^3 G_4} \frac{\delta_{\alpha\beta}}{|\vec{x}_{12} |^2} = \frac{\sqrt2 N^{3/2} k^{1/2}}{3\pi^3} \frac{\delta_{\alpha\beta}}{|\vec{x}_{12} |^2} \,, \\
\<\co_1(\vec x_1)\co_2(\vec x_2)\co_3(\vec x_3)\> &=& \frac{L^2}{4\pi^4G_4}\frac{1}{|\vec x_{12}|  |\vec x_{23}| |\vec x_{31}|}
   = \frac{\sqrt2 N^{3/2} k^{1/2}}{6\pi^4}\frac{1}{|\vec x_{12}|  |\vec x_{23}| |\vec x_{31}|} \,.
 \eea
In these expressions, $ \vec{x}_{ij} \equiv \vec{x}_i - \vec{x}_j$.  The right side is the result of supersymmetric localization applied to the ABJM theory with gauge group $U(N)_k\times U(N)_{-k}$ and Chern-Simons level $k = 1$ or $2$.   The left side comes from the AdS/CFT dual that we have been discussing.
Equality of the  coefficients follows from the AdS/CFT dictionary where it is deduced from the properties of $M2$ branes in 11-dimensional supergravity.   

We would like to wind up our discussion with two remarks:\\
i.  One important point  concerns alternate quantization.  In a supersymmetric bulk theory, one must identify the sources of all elementary fields in the model in terms of the asymptotic coefficients in the large $r$ behavior of solutions of the equations of motion. The sources  are functions on the boundary that transform among themselves under global supersymmetry rules determined from the large $r$ limit of the bulk transformation rules. The complete Legendre transform is a functional of the sources, and it must be supersymmetric.\\
ii.  It was to simplify the discussion that the basic ideas were outlined for truncations of the complete $\cn =8,~D=4$ gauged supergravity.  All results do extend to the complete theory.\\
For further information on these points and to clarify earlier parts of this section, we invite readers to consult \cite{FPPW}. 

\section{Acknowledgments}
The author thanks Prof. A. Zichichi for the opportunity to present these ideas and his warm hospitality.  He is supported in part by the US National Science Foundation  Grant No. PHY 1620045 and as Templeton Visiting Professor at Stanford.

\end{document}